\documentclass[aps,prl,twocolumn,superscriptaddress,showpacs,tighten]{revtex4}

\usepackage[dvips]{color} 
\usepackage{graphicx}
\usepackage{bm}      
\usepackage{latexsym}
\usepackage{amsmath,amssymb}
\usepackage{color}
\usepackage{stackrel}
\usepackage{accents}

\DeclareMathOperator{\sgn}{sgn}
\newcommand{\ord}[1]{\bm{\mathit{O}}\left(#1\right)}

\newcommand{\intl}[1]{\int\limits_{#1}}
\newcommand{\vex}[1]{\bm{\mathrm{#1}}}

\newcommand{\ts}[1]{\textstyle{#1}}

\newcommand{\bsub}{\begin{subequations}}
\newcommand{\esub}{\end{subequations}}

\newcommand{\T}{\mathsf{T}}

\newcommand{\Nabla}{\bm{\nabla}}

\newcommand{\kap}{\kappa}

\newcommand{\sigh}{\hat{\sigma}}
\newcommand{\sigb}{\hat{\bm\sigma}}

\newcommand{\kah}{\hat{\kappa}}

\newcommand{\muh}{\hat{s}}

\newcommand{\lf}{\mathcal{L}}
\newcommand{\rt}{\mathcal{R}}
\newcommand{\msf}[1]{\mathsf{#1}}
\newcommand{\LF}{\mathsf{L}}
\newcommand{\RT}{\mathsf{R}}

\newcommand{\tk}{\hat{\mathfrak{t}}_{\kappa}}
\newcommand{\tmuR}{\hat{\mathfrak{t}}_{S R}}

\newcommand{\gamr}{\gamma_{\msf{r}}}
\newcommand{\gami}{\gamma_{\msf{i}}}
\newcommand{\Proj}{\hat{\mathsf{P}}}

\begin{document}

\title{Interaction-mediated surface state instability in disordered three-dimensional topological superconductors with spin $SU(2)$ symmetry}
\author{Matthew S.\ Foster}
\email{matthew.foster@rice.edu} 
\affiliation{Department of Physics and Astronomy, 
	     Rice University, 
             Houston, 
             Texas 77005,
	     USA}
\author{Emil A.\ Yuzbashyan}
\affiliation{Center for Materials Theory, Department of Physics and Astronomy, 
	     Rutgers University, 
	     Piscataway, 
	     New Jersey 08854, 
	     USA}
\date{\today}

\begin{abstract}
We show that arbitrarily weak interparticle interactions destabilize the surface states of 
3D topological superconductors with spin $SU(2)$ invariance (symmetry class CI), in the 
presence of non-magnetic disorder. The conduit for the instability is disorder-induced wavefunction 
multifractality. We argue that time-reversal symmetry breaks spontaneously at the surface, so 
that topologically-protected states do not exist for this class. The interaction-stabilized 
surface phase is expected to exhibit ferromagnetic order, or to reside in an insulating plateau 
of the spin quantum Hall effect.
\end{abstract}


\maketitle


The existence of novel delocalized surface states is a key signature
of 3D topological phases of matter \cite{TISC,TopClassesDirty,TopClassesSC}. 
These states envelop a fully-gapped, yet topologically ``twisted'' bulk and can display 
exceptional properties such as the quantized magnetoelectric effect and Majorana fermions \cite{TISC}.
A complete classification \cite{TopClassesDirty,TopClassesSC} for (effectively) non-interacting 
particles has demonstrated that only five classes of topological phases and associated surface states
arise in 3D.

An important development
\cite{TopClassesDirty} has 
been the incorporation of disorder effects on 2D surface states. This is crucial 
because the terminating facets of a bulk 3D crystal inevitably host structural defects 
and impurities. 
The topologically nontrivial bulk is linked \cite{TopClassesDirty} to an effective 
low-energy surface theory of 2D Dirac fermions, 
perturbed by random impurity potentials \cite{BL,Loc}.  
Each of the
five classes of 3D topological phases 
is
``protected'' from the effects of time-reversal invariant (i.e., non-magnetic) 
disorder, in the sense that 
at least one surface Dirac wavefunction escapes 
Anderson localization \cite{TopClassesDirty,Loc}. 

Unlike uniform plane waves, the extended 2D energy eigenstates enveloping
a surface-disordered topological phase exhibit wild spatial amplitude fluctuations. 
These arise from quantum interference due to multiple impurity scattering, and
manifest in the local density of states (LDOS).
The pattern of LDOS fluctuations presents an intricate structure, characterized 
by an infinite set of scaling dimensions associated to interwoven fractal measures of the 
surface, a feature known as \emph{multifractality} \cite{Loc}. 
The evasion of localization in favor of multifractal scaling is rare in 2D, 
and is a signature of topological protection in the presence of disorder \cite{MFC-TI}.

In this Letter, we show that topological protection can be undermined by 
interparticle interactions. In particular, we study the \emph{combined} effects of 
multifractal LDOS fluctuations
\emph{and} 
interactions
upon the surface Andreev bound states of 3D topological superconductors.
Because the bulk condensate screens the long-ranged Coulomb force, surface quasiparticles
interact only via short-ranged potentials. In the clean limit, the vanishing density of
states for the 2D Dirac surface band implies that weak short-ranged interactions are 
irrelevant, i.e.\ the surface states remain ``protected.'' 
However, it is known that
disorder-induced LDOS 
multifractality can amplify interaction effects, such as 
pairing correlations near the superconductor-insulator transition \cite{SCMFC}.
With physics dominated by its 
surface, the complete picture of a 3D topological phase must incorporate
\emph{both} disorder-induced quantum interference and interactions \cite{OstrovskyGornyiMirlin10}.

\begin{figure}[b!]
   \includegraphics[width=0.4\textwidth]{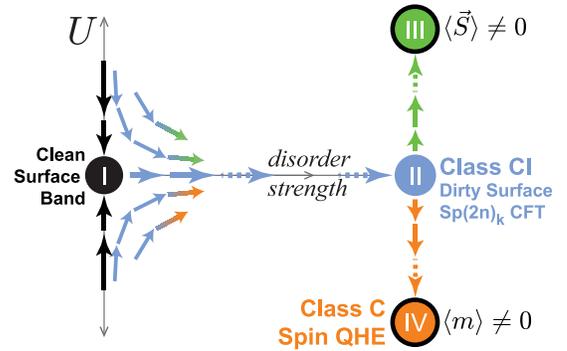}
   \caption{
	Phase portrait sketch for the surface physics 
	of a 3D time-reversal invariant, spin $SU(2)$ symmetric 
	topological superconductor. 
	The vertical axis is the interaction 
	strength $U$ [Eq.~(\ref{HI})], while the horizontal axis measures
	non-magnetic disorder. 
	Although the non-interacting system has a disorder-stabilized
	phase with delocalized (``protected'') surface states {\bf (II)},
	it is destroyed by arbitrarily weak interactions [Eq.~(\ref{UFlow})].
	Instead, at zero temperature, we expect that the surface exhibits 
	broken spin symmetry [$U > 0$ $\Rightarrow$ {\bf (III)}],
	or the spin quantum Hall effect [$U < 0$ $\Rightarrow$ {\bf (IV)}].
	In either scenario, interactions break time-reversal symmetry spontaneously. 
	}
   \label{PhaseDiag}
\end{figure}

Specifically, we demonstrate that arbitrarily weak interactions 
(consistent with bulk symmetries)
destabilize the non-interacting 
surface states of 3D topological superconductors with spin $SU(2)$ symmetry (class CI \cite{TopClassesDirty,TopClassesSC}), 
in the presence of non-magnetic disorder. 
Multifractal LDOS fluctuations
enhance the interactions, facilitating the instability.
We argue that 
time-reversal symmetry breaks spontaneously at the surface, 
so that ``protected''
surface states \emph{do not exist} in this class. Depending upon the sign of the relevant
interaction coupling $U$ [see Fig.~\ref{PhaseDiag}, Eqs.~(\ref{HI}) and (\ref{UFlow})], 
the surface should develop ferromagnetic order, or enter an insulating plateau state 
of the spin quantum Hall effect \cite{SpinQHE}. 
Our result provides impetus to identify a suitable material for the class CI bulk
as an avenue to realize the spin quantum Hall phase. 
A similar analysis for the 3D topological superconductor class AIII will be published 
elsewhere \cite{FosterUnpub}.


Three of the five 3D topological symmetry classes can be realized as time-reversal invariant
superconductors, distinguished by the degree of electronic spin conservation. 
In a 3D topological superconductor, Cooper pairing leads to a fully-gapped
quasiparticle band in the bulk, associated to an integer-valued winding number 
$\nu$ \cite{TopClassesDirty,TopClassesSC}. The modulus $|\nu|$ equals 
the number of flavors (or ``valleys'') of 2D
quasiparticle bands that appear at the sample surface, with energies that infiltrate the bulk gap. 
In the clean limit, the surface states exhibit a massless Dirac character at low energies; 
the Dirac point appears precisely at the bulk chemical potential (inside the gap) due to particle-hole symmetry.

In this paper, we study a universal low-energy model for the 2D surface states of a 3D class CI 
topological superconductor. In contrast to the spin-orbit-coupled $\mathbb{Z}_2$ topological insulators,
a CI superconductor has full spin $SU(2)$ symmetry. The non-trivial topology 
arises through the entwining of orbital degrees of freedom, including 
non-(simple) s-wave pairing \cite{SRL2009,ReadGreen2000}. For class CI $\nu$ is 
even because Dirac surface bands appear in time-reversal conjugate pairs \cite{TopClassesDirty,SRL2009,Supinfo}. 
We consider the generic case with $|\nu| = 2 k$, $k \in \{1,2,3,\ldots\}$.
Neglecting interactions, the Hamiltonian for the surface theory is 
\begin{align}\label{H_D}
	H_D
	=&
	\int
	d^2 \vex{r} \,
	\psi^\dagger
	\left\{
	-	
	\sigb \cdot
	\left[
	i \Nabla
	-
	\vex{A}_i(\vex{r})
	\tk^i
	\right]
	\right\}
	\psi.
\end{align}
The fermion field $\psi$ is a complex Dirac spinor
with pseudospin $\sigma \in \{1,2\}$ and valley $v \in \{1,\ldots,2k\}$ indices, i.e.\
$\psi \rightarrow \psi_{\sigma,v}$ when all indices
are displayed. 
The pseudospin components $\psi_{1,v}$ and $\psi_{2,v}$ 
are linear combinations of the Nambu elements 
$c_{\uparrow,v,\lambda}$ and $c^\dagger_{\downarrow,\bar{v},\lambda'}$.
These annihilate (create) spin up (down) electrons in valley
$v$ ($\bar{v}$). (Under time-reversal, $v$ and $\bar{v}$ interchange.)
The indices $\{\lambda,\lambda'\}$ label additional orbital 
(e.g.\ sublattice) degrees of freedom, whose precise interpretation 
depends upon bulk microscopics. A 3D class CI lattice model with $\nu = \pm 2$ 
appeared in Ref.~\cite{SRL2009}.

For a 3D topological superconductor, a key consequence of the non-trivial bulk is the 
special form that time-reversal symmetry adopts on the surface. 
If we write $H_D \equiv \psi^\dagger \hat{h} \psi$, with $\hat{h}$ the single-particle
Hamiltonian operator, then the usual time reversal symmetry for spin-$1/2$ electrons
in the bulk translates into the following \emph{chiral} condition on the surface 
\cite{TopClassesDirty,SRL2009,FosterUnpub,Supinfo}:
\begin{align}\label{tri}
	- \hat{\sigma}^3 \, \hat{h} \, \sigma^3 = \hat{h}.
\end{align}
This implies that any surface disorder that does not break time-reversal
(including non-magnetic impurities)
can manifest only as a random vector potential in the low-energy Dirac description. 
[Recall that $\psi$ in Eq.~(\ref{H_D}) carries $U(1)$ spin, rather 
than electric charge. In this language, vector potentials couple to 
time-reversal invariant spin and valley currents.]
A homogeneous perturbation such as a chemical potential shift, or a time-reversal 
invariant pairing of the \emph{surface} quasiparticles can be eliminated 
by a gauge transformation. Moreover, an energy gap (Dirac mass term) cannot
appear at the surface of a topological superconductor unless time-reversal is broken. 
For class CI, non-magnetic disorder induces scattering between the $2 k$ valleys, 
in the form of the non-abelian valley vector potential $\vex{A}_i(\vex{r}) \, \tk^i$ in
Eq.~(\ref{H_D}). Here $\tk^i$ denotes a $2 k \times 2 k$ generator of the 
group $Sp(2k)$. (The group is symplectic due to the spin symmetry
\cite{Supinfo}.) In the absence of interactions, elastic scattering 
due to vector potential disorder begets delocalized, multifractal surface states,
many properties of which can be computed exactly via conformal field theory (CFT) 
\cite{BL,NersesyanTsvelikWenger94,MudryChamonWen96}. 

We first consider the effects of disorder upon the non-interacting surface states. 
Below we describe the physics and main idea of the CFT method.
A technical summary can be found in Ref.~\cite{Supinfo},
while a more comprehensive discussion will appear elsewhere \cite{FosterUnpub}. 
The spatial character of the surface state wavefunctions (localized versus extended)
can be ascertained via disorder-averaged moments of physical observables, such as the 
conductance or the local density of states (LDOS). To facilitate this, we replicate 
$\psi_{\sigma,v} \rightarrow \psi_{\sigma,v,a}$, where the replica index $a \in \{1,\ldots,n\}$, 
and we are to take $n \rightarrow 0$ at the end of the calculation \cite{Loc}. 
Symmetry is the primary tool employed in the following, so we will rewrite Eq.~(\ref{H_D}) 
in a manifestly symmetric form.
We decompose $\psi$ and $\psi^\dagger$ into ``left'' $\lf$ and ``right'' 
$\rt$ fields,
\begin{align}
	\left\{\lf_{\uparrow,v,a},\lf_{\downarrow,v,a}\right\}
	\equiv&
	\left\{
	\psi_{1,v,a},\,
	\psi_{2,v',a}^\dagger (\kah^2)_{v',v} 
	\right\},\nonumber\\
	\left\{\rt_{\uparrow,v,a},\rt_{\downarrow,v,a}\right\} 
	\equiv&
	\left\{
	\psi_{2,v,a},\,
	\psi_{1,v',a}^\dagger (\kah^2)_{v',v} 
	\right\}.
\end{align}
Here and below, repeated indices are summed.
$\lf_{s,v,a}$ denotes a $4 n k$-component spinor;
the index $s$ ($v$) transforms in the fundamental representation of the
spin $SU(2)$ [valley $Sp(2 k)$] symmetry.
We also define
\begin{equation}\label{LFRTDef}
	\LF \equiv \lf^\T i \muh^2 \kah^2 \rightarrow \LF_a^{s,v},\;
	\RT \equiv \rt^\T i \muh^2 \kah^2 \rightarrow \RT_a^{s,v}. 
\end{equation}
$\LF_a^{s,v}$ and $\RT_a^{s,v}$ transform in the conjugate representations 
of the spin and valley symmetry groups; $\muh^2$ and $\kah^2$ are spin and valley
antisymmetric Pauli matrices \cite{footnote--Invts}.
Eq.~(\ref{H_D}) can be rewritten as $H_D = H_0 + \delta H_D$,
where
\begin{align}
	H_0 
	=& 
	\,
	i
	\int d^2\vex{r}
	\left[
	\LF \, \bar{\partial} \, \lf
	+ 
	\RT \, \partial \, \rt
	\right],
	\label{H0}
	\\
	\delta H_D
	=&
	\int d^2\vex{r}
	\left[
	J_{\kappa}^i \bar{A}_i + \bar{J}^i_{\kappa} A_i
	\right].
	\label{dH_D}
\end{align}
In Eq.~(\ref{H0}), we have switched to complex spatial coordinates $\{z,\bar{z}\} = x \pm i y$,
$\{\partial,\bar{\partial}\} \equiv \frac{1}{2}(\partial_x \mp i \partial_y)$. The valley disorder appears
in Eq.~(\ref{dH_D}), where 
$\{A_i,\bar{A}_i\} \equiv - i (A_i^x \mp i A_i^y)$. 
This couples to the valley $Sp(2 k)$ current, which has the holomorphic component 
$
	J^i_{\kap}
	\equiv
	- (i/2) \LF \, \tk^i \, \lf.
$

Eq.~(\ref{H0}) is manifestly invariant under chiral (independent left and right) 
spin $SU(2)$, valley $Sp(2k)$, and replica $SO(n)$ transformations. 
The symmetry group enlarges to $SO(4 n k)$ if we include operations that mix all three index types.
This free fermion theory is equivalent to the $SO(4 n k)_1$ Kac-Moody CFT \cite{CFT}.
The latter has a special property known as a conformal embedding rule \cite{CFT,Affine,footnote--confembed}, 
which gives a decomposition into a ``product'' of two other CFTs: $Sp(2 n)_k$, associated to the 
(spin)$\times$(replica) 
invariance of Eq.~(\ref{H0}), and $Sp(2 k)_n$, associated to the valley symmetry. 

The delocalization physics of the non-interacting surface with Hamiltonian $H_0 + \delta H_D$
is the same as in Refs.~\cite{NersesyanTsvelikWenger94,MudryChamonWen96},
which dealt with 2D Dirac fermions coupled to a random $SU(N)$ vector potential. 
Disorder is a relevant perturbation to the clean fermion theory \cite{NersesyanTsvelikWenger94}.
Crucially, the impurity potential couples only to the valley current $J^i_{\kap}$ in Eq.~(\ref{dH_D}).
This leads to a ``fractionalization'' of the original $SO(4 n k)_1$ CFT: the valley $Sp(2 k)_n$
sector localizes, leaving behind the ``critical'' (delocalized) spin-replica $Sp(2n)_k$ sector 
\cite{NersesyanTsvelikWenger94}. The latter is used to compute the scaling behavior 
disorder-averaged observables. 
Even in the absence of interactions, disorder localizes all surface states away 
from zero energy \cite{footnote--NonZeroNRG}; this is different from the case
of a single Dirac fermion on the surface of a 3D topological insulator \cite{3DTILoc}.
However, the localization length diverges upon approaching the chemical potential,
so that the zero energy state at the Dirac point remains completely 
delocalized (``topologically protected'').

The disorder-induced spatial fluctuations of the LDOS $\nu(\varepsilon,\vex{r})$ 
are encoded in the \emph{multifractal spectrum} $\tau(q)$ \cite{Loc,MFC-TI}. 
The $\tau(q)$ spectrum measures the sensitivity of extended wavefunctions to the sample 
boundary. 
A large $L\times L$ area of the surface is finely partitioned
into a grid of boxes of size $a \ll L$. One then defines the
box probability $\mu_n$ and inverse participation ratio 
$\mathcal{P}_{q}$,
\begin{align}\label{Box Prob}
	\mu_n(\varepsilon) 
	\equiv
	\intl{\mathcal{A}_n}
	d^{2}\vex{r} \,
	\nu(\varepsilon,\vex{r}),
	\;\;\;
	\mathcal{P}_{q}(\varepsilon)
	\equiv
	\sum_{n} \left[\frac{\mu_n(\varepsilon)}{\bar{\nu}}\right]^q,
\end{align}
where $\mathcal{A}_n$ denotes the $n^{\text{th}}$ box
and $\bar{\nu} = \sum_{i} \mu_i$ is the global DOS. 
When $\varepsilon$ is tuned to a critical delocalization energy 
(such as a mobility edge), $\mathcal{P}_{q} \sim (a/L)^{\tau(q)}$,
where the exponent $\tau(q)$ is both self-averaging and universal \cite{ChamonMudryWen96}.
The multifractal spectrum thus provides a unique fingerprint for 
spatial fluctuations in a particular symmetry class. 
In the field-theoretic description,
the $q^{\mathrm{th}}$ moment of the disorder-averaged LDOS ($q \in \{1,2,3\ldots\}$) 
is associated to a particular composite operator $\mathcal{O}_q$, with scaling dimension $\Delta_q$. 
The set of such dimensions determines the multifractal spectrum via
$\tau(q) = 2(q - 1) + \Delta_q - q \Delta_1$ \cite{Loc,MFC-TI}.
By contrast, localized states are insensitive to the sample boundary
for sufficiently large $L$ and have $\tau(q) = 0$.

For the class CI surface, we have identified the operators 
that represent disorder-averaged LDOS moments; 
these are a subset of the primary fields in the $Sp(2n)_k$ CFT.
As a result, we obtain the 
exact disorder-averaged multifractal spectrum at zero energy \cite{FosterUnpub,Supinfo},
\begin{equation}\label{MFCSpec}
	\tau(q) 
	=
	(q - 1)
	\left[2 - {{\frac{q}{2(k+1)}}}
	\right].
\end{equation}
For $k = 1$, Eq.~(\ref{MFCSpec}) agrees with previous calculations 
\cite{MudryChamonWen96}; the form for general $k$ is new.
One of the main results of this paper, Eq.~(\ref{MFCSpec}) proves that the 
non-interacting surface states at the bulk chemical potential remain delocalized,
a consequence of the bulk topological order.


Now we turn to interparticle interactions. Robust 
surface states must be protected from the \emph{combined} effects of 
both disorder and interactions.
In a weakly-interacting fermion gas, the low-energy behavior of the 
density of states completely determines the importance 
of short-ranged interactions.
The lowest-order (tree level) renormalization group (RG) equation for a generic four-fermion
coupling $U$ is \cite{FosterUnpub}
\begin{align}\label{UFlowFrame}
	d \ln U/d l
	=
	\Delta_1 - 
	\Delta_2^{(U)}
	+ \ord{U},
\end{align}
where $l$ denotes the log of the RG length scale such as the system size.
In a \emph{clean} 2D system, $\Delta_2^{(U)} = 2 \Delta_1$,
with $\Delta_1$ the scaling dimension of the LDOS. 
For the clean Dirac surface band, $\Delta_1 = 1$, so that 
weak short-ranged interactions are strongly irrelevant. By contrast, 
a negative $\Delta_1$ (due, e.g., to a van Hove singularity) would imply that $U$ is relevant,
signaling a potential instability.
With impurities present, the exponents $\Delta_1$ and $\Delta_2^{(U)}$ denote 
scaling dimensions of the disorder-averaged LDOS and four-fermion interaction, respectively.
The latter satisfies the lower bound $\Delta_2^{(U)} \geq \Delta_2$ \cite{FosterUnpub}, 
where $\Delta_2$ is the dimension of the second LDOS moment that 
determines $\tau(2)$.
The crucial point is that $\Delta_2$ is independent of, and \emph{strictly less than} $2 \Delta_1$ 
for a multifractal delocalized state in a disordered system \cite{Duplantier91}.
Impurity-mediated LDOS fluctuations can therefore amplify short-ranged interaction effects, 
by increasing the overlap of single particle wavefunctions in local regions.
This is particularly relevant for an interaction $U$ that saturates the bound 
$\Delta_2^{(U)} = \Delta_2 < 2 \Delta_1$.

Physically, we expect that the important interactions include a spin exchange channel 
(because spin is a conserved hydrodynamic mode) and a Cooper pairing interaction (because disorder 
respects time-reversal). The former is written $- \vec{S} \cdot \vec{S}$, where $\vec{S}$ denotes the 
spin density. As discussed below Eq.~(\ref{tri}), pairing of the surface quasiparticles does not open a gap 
unless time-reversal is simultaneously broken. The latter occurs when the Dirac mass operator 
$m \equiv \psi^\dagger \sigh^3 \psi$ develops an expectation value. 
This can be understood explicitly in the 3D CI topological superconductor lattice model of Ref.~\cite{SRL2009},
which features real d-wave pairing in the bulk. In that model, $m$ is interpreted as a 
sum of pairing operators: $ m \sim -i c^\dagger_{\uparrow} c^\dagger_{\downarrow} + i c_{\downarrow} c_{\uparrow} $,
where $c_{s}$ annihilates a lattice electron. Crucially, $m$ is odd under time-reversal, due 
to the factors of $i$. Thus, a non-zero expectation $\langle m \rangle \neq 0$ would imply (``$d + i s$'') pairing of 
the surface quasiparticles, opening an energy gap and breaking time-reversal symmetry.
The resulting state is an insulating plateau of the spin quantum Hall effect (see below). 
An attractive Cooper pairing interaction can be written as $- m^2$. 

To keep the analysis general, we enumerate all four-fermion interactions that 
preserve the bulk symmetries [time-reversal invariance, spin $SU(2)$, and 
valley $Sp(2k)$ symmetry]. This necessitates the incorporation of a third 
interaction channel $J^{\gamma}_{S} \bar{J}^{\gamma}_{S}$, where $J^{\gamma}_{S}$ is the holomorphic spin current.
The replicated interaction Hamiltonian for the CI surface is \cite{Supinfo}
\begin{align}\label{HI}
	H_I
	=&\,
	\sum_{a = 1}^n
	\int d^2\vex{r}
	\left[
	U 
	\left(
	m_a m_a - 4 \vec{S}_a \cdot \vec{S}_a
	\right)
	+
	V 	
	J^{\gamma}_{S a} \bar{J}^{\gamma}_{S a}
	\right.
	\nonumber\\
	&\,
	\left.
	+
	W
	\left(
	3 m_a m_a + 4 \vec{S}_a \cdot \vec{S}_a
	-
	\frac{1}{k} J^{\gamma}_{S a} \bar{J}^{\gamma}_{S a}
	\right)
	\right].
\end{align}
The interaction strengths $\{U,V,W\}$ are defined so as to couple to 
RG eigenoperators, in the presence of disorder.
In the minimal two valley realization ($k = 1$), the $W$-channel interaction does not exist.
For that case only, 
$J^{\gamma}_{S a} \bar{J}^{\gamma}_{S a} = 3 m_a m_a + 4 \vec{S}_a \cdot \vec{S}_a$.

Our task is to evaluate Eq.~(\ref{UFlowFrame}) in the disordered,
non-interacting CI surface theory for the three interaction operators
in Eq.~(\ref{HI}) \cite{IQHP}.
Using the $Sp(2n)_k$ CFT, 
we have found that one particular operator controls the scaling 
of \emph{both} the second LDOS moment and the interaction $U$, 
leading to $\Delta_2^{(U)} = \Delta_2 = 0$, while $\Delta_1 = 1/2(k+1)$ \cite{FosterUnpub,Supinfo}.  
The main result of this paper follows,
\begin{align}\label{UFlow}
	\frac{d U}{d l}
	=
	\frac{U}{2(k+1)} + 
	\ord{U^2},
\end{align}
which implies that the interaction $U$ in Eq.~(\ref{HI}) grows
at longer wavelengths, destabilizing the non-interacting, dirty surface, for 
any number of $2 k$ valleys.
By contrast, the other interactions $V$ and $W$ remain irrelevant for any $k$, satisfying
$
	\frac{d \ln V}{d l} = - \frac{4k + 3}{2(k+1)},\;
	\frac{d \ln W}{d l} = - \frac{3}{2(k+1)}	
$
\cite{FosterUnpub,Supinfo}.
We conclude that while weak interactions are suppressed 
in the clean limit by the vanishing density of states at the Dirac point, 
surface disorder strongly renormalizes the interaction channel $U$, making it relevant. 

Eq.~(\ref{UFlow}) can be understood as an enhancement of interaction matrix elements in 
the eigenbasis of the disordered theory: local accumulations of the DOS due to wavefunction
multifractality induce stronger interactions between the surface quasiparticles.
The amplification of the particular interaction channel $U$ over the others 
signals the instability of the non-interacting surface to spontaneous time-reversal symmetry 
breaking. 
From Eq.~(\ref{HI}), we anticipate (at least local) ferromagnetic order $\langle \vec{S} \rangle \neq 0$ 
when $U \rightarrow + \infty$. Without time-reversal symmetry, the surface is not ``topologically protected'' 
\cite{TISC,TopClassesDirty,TopClassesSC}, and we expect Anderson localization of all surface states \cite{TopClassesDirty,Loc}.
However, we cannot rule out an exotic metallic phase when spin symmetry is also broken \cite{Chalker01}.
By contrast, $U \rightarrow -\infty$ should cause Cooper pairing of 
the surface quasiparticles. Treating the relevant interaction in mean field theory,
one replaces $m^2 \rightarrow 2 \langle m \rangle \psi^\dagger \sigh^3 \psi$ in Eq.~(\ref{HI}).
A non-zero Dirac mass opens an energy gap, producing an insulating surface. 
Time-reversal symmetry is broken because $\langle m \rangle \neq 0$ implies surface pairing at a 
non-zero superfluid phase angle with respect to the bulk.

To lowest order in $(1/k)$, Eq.~(\ref{UFlow}) agrees with a perturbative result \cite{DellAnna} 
obtained using the non-linear sigma model \cite{Loc,Supinfo}. 
The calculations in Ref.~\cite{DellAnna} were performed in the context of a non-topological 2D
system of gapless superconductor quasiparticles, subject to disorder and interactions 
with spin $SU(2)$ symmetry and time-reversal invariance. The $Sp(2 n)_k$ CFT employed
here has a sigma model representation with the same structure,
but augmented with a Wess-Zumino-Witten (WZW) term \cite{CFT}. In the $k \gg 1$ limit, this
model becomes weakly-coupled, and the WZW term can be ignored. 
The results of Ref.~\cite{DellAnna} therefore provide a non-trivial check of our analysis 
in the many-valley limit. In addition, at one loop in the sigma model calculation,
RG flow equations beyond linear order in the interaction strengths can be obtained,
because the sigma model treats interactions non-perturbatively via RPA and BCS-type summations. 
For the 2D class CI quasiparticle system, the sigma model generically predicts an instability 
of the ``metallic'' phase signaled by the divergence of the spin exchange or BCS pairing interaction strengths 
\cite{DellAnna,Supinfo}. This provides evidence for the absence
of an interacting, time-reversal invariant fixed point.

The insulating state that occurs for $\langle m \rangle \neq 0$ 
preserves spin $SU(2)$ symmetry. This state resides in a plateau of
the so-called spin quantum Hall effect \cite{SpinQHE}, analogous to the
``half-integer'' quantum Hall phase at the surface of a 3D $\mathbb{Z}_2$
topological insulator with broken time-reversal symmetry \cite{TISC,SRL2009}. 
The quantized spin Hall conductance \cite{SpinQHE} is $\sigma^s_{x y} = \frac{1}{h}\left(\frac{\hbar}{2}\right)^2  p$,
with $p = k \sgn \langle m \rangle$ if valley symmetry is unbroken 
\emph{on average} (i.e., after disorder-averaging). If valley symmetry remains
broken even after disorder-averaging, then $p \in \{-k,-k+2,\ldots,k-2,k\}$;  
see also Ref.~\cite{SpinQHE}.
Our results are summarized in Fig.~\ref{PhaseDiag}.


In conclusion, we have demonstrated that interactions destabilize class CI 
disordered surface states in 3D.
We have argued that time-reversal symmetry breaks spontaneously, and that the 
CI topological superconductor surface enters into either a ferromagnetic or a spin quantum Hall phase.
These are expected to be interaction-stabilized Anderson insulators.
The other 3D topological superconductor classes AIII and DIII
also admit WZW CFT descriptions \cite{TopClassesDirty}. 
The minimal surface state (single Dirac valley) realization for each of these is stable against 
disorder and short-ranged interactions \cite{TopClassesDirty,FosterUnpub}. 
Results for class AIII with multiple valleys will appear elsewhere \cite{FosterUnpub},
while class DIII is an important topic for future work.

M.S.F.\ thanks Andreas Ludwig for helpful discussions. 
This work was supported by the NSF under Award No.~DMR-0547769 and by the David and Lucille 
Packard Foundation.

\newpage

\begin{widetext}
\section{Supplementary information}


In this supplement, we provide the minimal details necessary to reproduce the results in the 
main text; a comprehensive discussion and analysis of class CI and AIII topological
superconductors will appear elsewhere \cite{FosterUnpubSM}.
We first summarize the symmetry structure of the class CI surface state theory. 
Then we provide the derivation of the multifractal spectrum and interaction
scaling dimensions. We close with a review of the many-valley limit and the
relation to the non-linear sigma model.

\section{Symmetries of the surface state theory}

In Eq.~(1) of the main text, the Dirac fermion $\psi_{\sigma,v}(\vex{r})$ is a surface-confined
projection \cite{TISCSM,TopClassesDirtySM,SRL2009SM} of a bulk field $\Psi_{\sigma,v,\tau}(\vex{r},z)$ 
that lowers the spin angular momentum by one. Relative to the surface, the bulk fermion carries 
an additional index $\tau \in \{1,2\}$. 
The four $\sigma\otimes\tau$ components of $\Psi_{v}$ are linear combinations of electron operators 
$c_{\uparrow,v,\lambda}$ and $c^\dagger_{\downarrow,\bar{v},\lambda'}$, where $(v,\bar{v})$ denote
a pair of valleys related by time-reversal, and $\{\lambda,\lambda'\}$ run over additional orbital (e.g. sublattice) 
labels \cite{SRL2009SM,FosterUnpubSM}.

In the bulk, a physical time-reversal operation sends $i \rightarrow -i$ (complex conjugation) and
\[
	c_{\uparrow,v} \rightarrow - \hat{M} \, c_{\downarrow,\bar{v}}, 
	\;\;\;
	c_{\downarrow,v} \rightarrow \hat{M} \, c_{\uparrow,\bar{v}}.
\]
In this equation, $\hat{M} \rightarrow M_{\lambda, \lambda'}$ is a symmetric, unitary matrix in orbital labels. 
Time-reversal induces the transformation 
$\Psi \rightarrow - i \sigh^3 (2 \hat{\mathsf{P}} - \hat{1}) \Psi^\dagger$,
where $\Proj = (\hat{1} + \hat{n}\cdot\hat{\vec{\tau}})/2$ projects onto a certain $\tau$-spin direction
$\hat{n}$. (The latter is basis-dependent and determined by microscopics \cite{FosterUnpubSM}.)
At the surface, the $\tau$-spin becomes ``locked,'' with $\psi(\vex{r}) \sim \Proj \Psi(\vex{r},z = 0)$ \cite{TopClassesDirtySM,SRL2009SM}.
This type of projection always occurs at the surface of a $d$-dimensional topological insulator or superconductor,
producing an anomalous state that is ``half'' of a normal $(d-1)$-dimensional system.

For the surface theory in Eq.~(1), time-reversal therefore appears as the antiunitary 
transformation \cite{TopClassesDirtySM,SRL2009SM,BLSM}
\begin{align*}
	\psi \rightarrow - i \sigh^3 \left[\psi^\dagger\right]^\T,\;\;\; i \rightarrow - i.
	\tag{S1}
\end{align*}
Spin $SU(2)$ symmetry requires invariance under $U(1)$ $\psi \rightarrow e^{i \theta} \psi$ 
and particle-hole $\psi \rightarrow i \sigh^1 \kah^2 \left[\psi^\dagger\right]^\T$
transformations; the latter encodes a $\pi$ rotation by $S^x$. 
Imposing these upon Eq.~(1) restricts the disorder to the $Sp(2 k)$ 
valley vector potential $\vex{A}_i \tk^i$, where $- \kah^2 \left[\tk^i\right]^\T \kah^2 = \tk^i$.
In these equations, $\kah^2$ denotes the valley $Sp(2k)$ invariant tensor
(antisymmetric $2k \times 2k$ block Pauli matrix).


\section{Conformal embedding}

The disordered, non-interacting class CI surface state theory is described by the 
$Sp(2 n)_k$ Kac-Moody CFT. To see this, we note that the free fermion Hamiltonian in Eq.~(5) 
of the main text is equivalent to the $SO(4 n k)_1$ current algebra \cite{CFTSM}. 
(Since we are discussing non-interacting fermions at this stage, we can trade the 2D Hamiltonian
for a 2+0-D Grassmann field action.)
The conformal embedding $SO(4 n k)_1 \supset Sp(2 n)_k \oplus Sp(2 k)_n$ \cite{AffineSM} implies
that the associated stress tensor $T(z)$ has the Sugawara decomposition \cite{CFTSM} 
\begin{align}\label{ConfEmb}
	T(z)
	= 
	{\ts{\frac{1}{2(n + k + 1)}}}
	\left[
	: J_{\kappa}^i J_{\kappa}^i :(z)
	\,+
	: J_{S R}^\alpha J_{S R}^\alpha :(z)
	\right],
	\tag{S2}
\end{align}
where  
\begin{equation}\label{KMCurrents}
	J^i_{\kap}
	\equiv
	{\ts{\frac{i}{2}}}
	\sum_{a}
	J^i_{\kap a},
	\;\;
	J^i_{\kap a}
	\equiv
	- \LF_a \, \tk^i \, \lf_a,
	\;\;
	J^{\alpha}_{S R}
	\equiv
	-
	{\ts{\frac{i}{2}}}
	\LF \, \tmuR^\alpha \, \lf. 
	\tag{S3}
\end{equation}
The replica-summed valley current $J^i_{\kap}(z)$ satisfies the $Sp(2 k)_n$ algebra. 
In Eqs.~(\ref{ConfEmb}) and (\ref{KMCurrents}), $J^{\alpha}_{S R}$ generates $Sp(2 n)$ spin $\times$ replica
space transformations; 
$( \tmuR^\alpha )_{s}{}^{s'}{}_{a a'}$ is a suitable
$2 n \times 2 n$ matrix. 
The current $J^{\alpha}_{S R}(z)$ satisfies the $Sp(2 n)_k$ algebra. 

The impurity potential $\{A_i,\bar{A}_i\}$ in Eq.~(6) of the text couples only to the valley 
Kac-Moody current $\{J^i_{\kap},\bar{J}^i_{\kap}\}$. As the disorder renormalizes towards strong coupling, it 
localizes the valley $Sp(2k)_n$ sector.
To obtain Eqs.~(8) and (11), we express the associated operators in terms of $Sp(2n)_k$
primary fields and descendants. The most relevant components govern the leading scaling behavior.


\section{Multifractal spectrum}

We consider first the $\tau(q)$ spectrum at level $k = 1$.
Primary fields are labeled by the $Sp(2 n)$ fundamental weights 
$\{\Lambda_{q}\}$, 
$q \in \{1,\ldots,n\}$ \cite{CFTSM}.
Weight $\Lambda_q$ 
corresponds to a rank $q$, fully antisymmetric tensor $\Omega^{(q)}_{[i_1 i_2 \cdots i_q]}$,
satisfying the traceless condition $(\muh^2)^{i j}\Omega_{[i j \cdots i_q]} = 0$.
Here, $i = \{s,a\}$ is a product of spin $s$ and replica $a$ indices; $(\muh^2)^{i j}$ is
the $Sp(2n)$ invariant tensor.
In the holomorphic sector, the $q^{\mathsf{th}}$ LDOS moment must correspond
to a tensor with $q$ distinct replica indices. We 
therefore associate 
\begin{align}\label{LDOSqOp}
	&
	\overline{(\psi_1^\dagger \psi_1) (\psi_2^\dagger \psi_2) \times \cdots \times (\psi_q^\dagger \psi_q)(\vex{r})}
	\nonumber\\
	&\;\;\;\;
	\Leftrightarrow
	\Omega^{(q)}_{[\{s_1,1\}\{s_2,2\} \cdots \{s_q,q\}]}(z) \, \bar{\Omega}^{(q)}_{[\{s_1',1\}\{s_2',2\} \cdots \{s_q',q\}]}(\bar{z})
	\nonumber\\
	&\;\;\;\;\;
	\phantom{\Leftrightarrow}	
	\times(i \muh^2 \muh^{\msf{3}})^{s_1,s_1'} (i \muh^2 \muh^{\msf{3}})^{s_2,s_2'} \times \cdots \times (i \muh^2 \muh^{\msf{3}})^{s_q,s_q'}.
	\tag{S4}
\end{align}
The left-hand side is the disorder-averaged $q^{\mathsf{th}}$ 
moment. The
right-hand side is a diagonal primary field labeled by 
$\Lambda_q$.
Here we have written each $Sp(2n)$ index as 
the spin-replica product
$\{s,a\}$; 
$\psi^\dagger \psi \rightarrow S^{3}$ is the
\emph{spin-projected} LDOS \cite{MFC-TISM} for the surface state. 
We find that
the most relevant contribution to the $q^{\mathsf{th}}$ LDOS moment is associated to the same 
$Sp(2n)$ representation $\Lambda_q$, for any $k \geq 1$.
The scaling dimension of the $q^{\mathsf{th}}$ 
moment is then given by \cite{CFTSM}
\begin{align}\label{Deltaq}
	\Delta_q = (2 q - q^2)/2(k + 1)
	\tag{S5}
\end{align}
in
the replica limit \cite{LocSM} $n \rightarrow 0$. 
Eq.~(8) follows.


\section{Interaction dimensions}

The flow equations for each of the three interaction couplings $U$, $V$, and $W$ in Eq.~(10) 
of the main text appear as in Eq.~(9), with $\Delta_1$ given as above
and $\Delta_{2}^{(U,V,W)}$ the scaling dimension of the associated field. 
We rewrite Eq.~(10) exploiting $Sp(2n)$ \cite{footnote--FierzSM} and $SU(2)$ Fierz identities:
\begin{align}
	\label{HISM}
	H_I
	=&
	\int d^2\vex{r}
	\left[
	2 U J_{\kappa a}^i \bar{J}_{\kappa a}^i
	+
	V J^{\gamma}_{S a} \bar{J}^{\gamma}_{S a}
	+
	2 W
	I_{v' a}^{v \gamma} 
	\bar{I}_{v a}^{v' \gamma}
	\right],
	\tag{S6}
\end{align}
where
$
	J^\gamma_{S a} \equiv -\LF_a \muh^\gamma \lf_a
$
denotes a replica-resolved spin current (no sum on $a$ is implied).
Here $\muh^{\gamma}$ denotes a spin space Pauli matrix. 
The holomorphic half of the last term in Eq.~(\ref{HISM}) is
$I_{v' a}^{v \gamma} \equiv \LF_{a}^{v} \muh^{\gamma} \lf_{v' a} + \delta^{v}_{v'} \frac{1}{2k} J^{\gamma}_{S a}$.
If we assume that the disorder-averaged theory is invariant 
under both spin $SU(2)$ and valley $Sp(2k)$ transformations, then 
the three channels $U$, $V$, and $W$ in Eq.~(\ref{HISM}) exhaust the 
possibilities for four-fermion interactions.

As $V$ couples to a KM current-current perturbation, $\Delta_{2}^{(V)} = 2$ 
\cite{CFTSM}.
The $U$ interaction involves the valley current $J^i_{\kappa a}$. 
This is \emph{not} a valley KM current, but rather a product of $Sp(2n)_k$ and 
$Sp(2k)_n$ primary fields. 
[Replica-resolved components cannot be extracted from the KM current $J^i_{\kappa}$ in Eq.~(\ref{KMCurrents}).]
In the $Sp(2n)_k$ theory, $J^i_{\kappa a}$
corresponds to a second rank tensor with equal replica indices, antisymmetrized
over spin (to obtain a singlet). The only choice is 
$
	J^i_{\kappa a}(z) \Leftrightarrow \Omega^{(2)}_{[\{s,a\} \{s',a\}]}(z),
$
i.e.\ the same representation 
$\Lambda_2$ 
that determines the scaling of the second
LDOS moment [Eqs.~(\ref{LDOSqOp}) and (\ref{Deltaq}) with $q = 2$]. 
As a result,
$\Delta_{2}^{(U)} = \Delta_2 = 0$.
Finally, the $W$ interaction operator $I_{v' a}^{v \gamma}$ 
is
a second rank tensor field with equal replica indices, symmetric in spin. 
The only choice has weight 
$2 \Lambda_1$, 
leading to $\Delta_{2}^{(W)} = 2/(k + 1)$ 
\cite{CFTSM}.
Via Eq.~(9), we obtain Eq.~(11) and the equations for $V$ and $W$ quoted in the text.


\section{Many valley (large $k$) limit: Review of the class CI Finkel'stein NL$\sigma$M results}

In Ref.~\cite{DellAnnaSM}, the effects of interactions upon gapless quasiparticles in disordered, 
non-topological 2D superconductors were considered using the Finkel'stein non-linear sigma model
framework \cite{BKSM}. 
In particular, the author studied 
symmetry class CI, appropriate to a spin singlet superconductor possessing spin $SU(2)$ symmetry
and time-reversal invariance in every realization of the disorder. 
The 2D topological surface state $Sp(2n)_k$ CFT studied in the present paper also possesses a sigma model description
with the same structure, but augmented by a WZW term \cite{TopClassesDirtySM,BLSM,NersesyanTsvelikWenger94SM,LocSM}.  
In the limit of many valleys $k \gg 1$, the WZW term can be ignored in the first approximation,
yielding a ``metallic phase'' with a large spin conductance proportional to $k$, independent of the disorder \cite{LocSM}. 
The advantage of working directly in the sigma model framework is that interactions are treated to all orders,
via RPA and BCS-type resummations. The disadvantage is that the sigma model becomes strongly coupled
for small $k \in \{1,2,3,\ldots\}$, so that the WZW term cannot be ignored and 
perturbation theory becomes unreliable.

The only interactions that appear in the CI sigma model treated in Ref.~\cite{DellAnnaSM} are
short-ranged, and include a spin triplet exchange coupling $\gamma_t$ and a Cooper pairing 
interaction $\gamma_c$. Here, we define $\gamma_{t,c} > 0$ for repulsive interactions in each channel;
these are dimensionless couplings in the limit of vanishing disorder strength.
Other (e.g.\ electric charge density-density) interaction channels do not couple to conserved hydrodynamic
diffusion modes, and are strongly irrelevant in the sigma model framework.

The one-loop RG equations for $\gamma_{t,c}$ are \cite{DellAnnaSM}
\begin{align*}
	\label{NLsMFull}
	\tag{S7}
	\begin{aligned}
	\frac{d \gamma_t}{d l}
	=&\,
	-
	\frac{\lambda}{2} \gamma_c
	\left(1 - \gamma_t\right)
	\left(1 - 2 \gamma_t\right),
	\\
	\frac{d \gamma_c}{d l}
	=&\,
	\frac{\lambda}{2}
	\left\{
	-3 \gamma_t - 2 \gamma_c
	+
	3 \gamma_c
	\left[
	\log(1 - \gamma_t) + \gamma_t
	\right]
	\right\}
	-
	\gamma_c^2.
	\end{aligned}
\end{align*}
In these equations, $\lambda$ is proportional to the 
dimensionless inverse spin conductance.
For weak interactions, these equations can be linearized in the
coupling strengths. The result is 
\begin{align*}\label{NLsM}
	\frac{d \ln \gamr}{d l}
	=
	\frac{\lambda}{2},
	\;\;\;
	\frac{d \ln \gami}{d l}
	=
	-\frac{3\lambda}{2},
	\tag{S8}
\end{align*}
where $\gamr \equiv \gamma_c - 3 \gamma_t$, and $\gami \equiv \gamma_c + \gamma_t$.
Since $\lambda \propto 1/k$ \cite{LocSM,CFTSM}, we find that
Eq.~(\ref{NLsM}) matches the RG equations for the relevant ($U \leftrightarrow \gamr$) and
irrelevant ($W \leftrightarrow \gami$) coupling strengths in the WZW surface state theory studied in the present paper,
Eq.~(11) and the text following, valid to lowest order in $1/k$.
The relevant sigma model coupling $\gamr$ is indeed the difference of repulsive singlet pairing and 
triplet exchange interactions, just as we found in the WZW model.

Integrating the full flow Eqs.~(\ref{NLsMFull}) numerically, one sees that the metallic phase is 
generically destroyed by one of two instabilities: either the triplet exchange interaction 
flows to minus infinity $\gamma_t \rightarrow - \infty$ (suggesting Stoner ferromagnetism), or 
the Cooper pairing strength flows to minus infinity $\gamma_c \rightarrow - \infty$, indicating a 
residual pairing instability for the gapless quasiparticles. We do not discuss here the back-reaction
of the interactions upon the spin conductance \cite{DellAnnaSM}, since the lowest order quantum corrections 
are modified by the WZW term.

\end{widetext}


\begin{thebibliography}{99}
\bibitem{footnote--Invts}
	In Eq.~(\ref{LFRTDef}), ``$\T$'' is the matrix transpose, while $\muh^2$ ($\kah^2$) 
	denotes the $2 \times 2$ ($2 k \times 2 k$) antisymmetric invariant tensor for 
	the spin $SU(2)$ [valley $Sp(2k)$].
\bibitem{footnote--NonZeroNRG}
	At non-zero energy, all states are localized. 
	Energy  breaks the 
	particle-hole
	symmetry, 
	altering the 
	class to AI. 
	This has no delocalized phase in 2D \cite{TopClassesDirty,Loc}.
\end{thebibliography}
\end{document}